\documentstyle[preprint,eqsecnum,aps]{revtex}

\begin{document}
\draft
%\preprint
\title{Theory of electron--hole asymmetry in doped {\em CuO$_2$} planes}
\author{R.J. Gooding and K.J.E. Vos}
\address{Dept. of Physics, Queen's University,\\
Kingston, Ontario CANADA K7L 3N6}
\author{P.W. Leung}
\address{Dept. of Physics, Hong Kong University of Science
and Technology,\\
Clear Water Bay, Hong Kong\\
\bigskip}
%\date{\today}
\date{To be published in Phys. Rev. B, Nov. 1, 1994.}
\maketitle
\bigskip
\bigskip
\bigskip
Keywords: Weakly doped high $T_c$ superconductors; $t - t^{\prime} - J$ model;
electronic \hfill\break ~~~~~phase separation.
\bigskip
\bigskip
\bigskip
%\newpage
\begin{abstract}
The magnetic phase diagrams, and other physical characteristics, of 
the hole--doped {\em La$_{2-x}$Sr$_x$CuO$_4$}
and electron--doped {\em Nd$_{2-x}$Ce$_x$CuO$_4$} high--temperature 
superconductors are profoundly different. Given that it is 
envisaged that the simplest Hamiltonians describing these systems 
are the same, viz. the $t - t^\prime - J$ model, this is surprising.
Here we relate these physical differences to their ground states'
single--hole quasiparticles, the spin distortions they produce,
and the spatial distribution of carriers for the multiply--doped systems.
As is well known, the low doping limit of the hole--doped material corresponds to
$\vec k = ({\pi\over 2}, {\pi\over 2})$ quasiparticles,
states that generate so--called Shraiman--Siggia long--ranged dipolar spin
distortions via backflow. These quasiparticles have been proposed to lead
to an incommensurate spiral phase, an unusual scaling of the magnetic susceptibility,
as well as the scaling of the correlation length defined by
$\xi^{-1} (x,T) = \xi^{-1} (x,0)~+~\xi^{-1} (0,T)$,
all consistent with experiment.  We suggest that
for the electron--doped materials the single--hole 
ground state corresponds to $\vec k = (\pi, 0)$ quasiparticles;
we show that the spin distortions generated by such carriers are short--ranged. 
Then, we demonstrate the effect of this single--carrier difference in many--carrier
ground states via exact diagonalization results by evaluating $S(\vec q)$ for 
up to 4 carriers in small clusters. Consistent with experiment, for the
hole--doped materials short--ranged incommensurate spin orderings
are induced, whereas
for the electron--doped system only commensurate spin correlations are found.
Further, we propose that there is an important difference between the spatial
distributions of mobile carriers for these two systems: for the hole--doped
material the quasiparticles tend to stay far apart from one another,
whereas for the electron--doped material we find tendencies consistent
with the clustering of carriers, and possibly of low--energy fluctuations
into an electronic phase separated state. Phase separation in this material
is consistent with the mid--gap states found by recent ARPES studies.
Lastly, we propose the extrapolation of an approach based on the
$t - t^\prime - J$ model to the hole--doped 123 system.

\end{abstract}


%\narrowtext
\newpage
\section{Introduction:}
\label{sec:intro}

The {\em CuO$_2$} plane based high--temperature superconductors
have anomalous normal state properties.  One part
of the normal state puzzle involves the spin orderings,
and for the majority of this paper we focus on the differing magnetic
phase diagrams for the hole--doped Bednorz--M\"uller {\em La$_{2-x}$Sr$_x$CuO$_4$}
compounds \cite {bednorz} in comparison to the electron--doped
{\em Nd$_{2-x}$Ce$_x$CuO$_4$} materials \cite {tokura}. In particular, for both of these
systems the $x=0$ phase possesses three--dimensional antiferromagnetic
long--ranged order \cite {birgeneau,electronmag}. However, one must heavily
dope the $Nd$ compound to destroy this
order, say $\sim$ 12 \%, while only a small doping of $\sim$ 2 \% is required to kill
long--ranged antiferromagnetic order in the $La$ cuprate material.
A schematic of the contrasting phase diagrams is given in Fig. 1. 

One important aspect of the magnetic orderings found in these systems
involves the kinds of spin correlations
that these systems exhibit when doped to levels greater than 2 \%. As mentioned above, for the
electron--doped material antiferromagnetic order persists until $\sim 12 \%$
doping levels are reached. However, when the hole--doped system has
say 7 \% holes in a {\em CuO$_2$} plane, short--ranged incommensurate
magnetic ordering is found. This appears experimentally in the dynamic magnetic response,
and has been found via inelastic neutron scattering for a number of 
intermediate doping levels \cite {aeppli,mason,thurston2}. Consequently,
an alternate way in which one can phrase the question concerning
the differing magnetic phase diagrams is: why does the electron--doped
system maintain commensurate antiferromagnetic order at doping levels
that induce incommensurate correlations in the hole--doped system?

Several theories have addressed the phase diagrams of these doped antiferromagnetic
insulators.
When the {\em La}-based system is doped, the carriers predominantly occupy
{\em O} sites, in particular as {\em O$^-$} states, whereas for the {\em Nd}-based 
material the carriers are believed to be associated with the addition of
an electron to a {\em Cu~3d~$^9$} ion, thus forming {\em Cu$^+$}. Then, a common explanation 
of these differing magnetic phase diagrams follows from the assumption of the complete 
localization of the carriers, and the ensuing spin distortions generated by 
such static defects. Namely:
(i) The dramatic reduction of 
the N\'eel temperature $T_N (x)$ in the hole--doped system results from frustrating 
ferromagnetic bonds being embedded in an antiferromagnetic background, the so--called Aharony
model \cite {aharony}. Such frustrated bonds generate a long--ranged spin distortion
with dipolar symmetry, and numerical studies \cite {polarons} have 
shown that a version of this perturbation renormalized by quantum fluctuations \cite{polarons,frenkel}
indeed alters the spin--spin correlation length in a fashion consistent with 
experiment \cite {keimer}. (ii) Fixed {\em Cu$^+$} sites in a background
antiferromagnetic lattice act like static vacancies, and subsequently
diminish the spin--wave stiffness \cite {manousakis} in a manner
consistent with experiment \cite {thurston1}. Monte Carlo studies
of this quantum dilution problem \cite {manousakis,behre} do not suffer from
the minus sign problem, and thus are capable of accurately characterizing the
spin correlation length. 

Unfortunately, these cannot be complete explanations of the differing magnetic phase
diagrams.  Near the antiferromagnetic phase boundary (temperature vs. doping)
the carriers are mobile in both of these systems \cite {batlogg,uji},
and thus one must come to an understanding of the differing reductions of 
the spin orderings for mobile not just localized carriers. This is one of the
focuses of this paper \cite {second}.

Here we put forward a proposal explaining the electron--hole asymmetry of the
magnetic phase diagrams shown in Fig. 1 for mobile carriers.
Our work builds on our earlier study \cite {GVL} that showed that
for the $t - J$ model and the hole--doped
system, a tendency towards short--ranged incommensurate
order vs. hole doping arises. The results of \cite {GVL} are consistent
with experiment \cite {aeppli,mason,thurston2}, and
in part is further verification of earlier exact diagonalization work of
Moreo et al. \cite {moreo1}, as well as Monte Carlo 
work \cite {moreo2,furukawa}. However, the important point of \cite {GVL} was that
this incommensurability only occurred for the single--hole
system having a ground state momentum of
$\vec k = \pm ({\pi\over 2}, \pm {\pi\over 2})$ 
(since $\vec k =({\pi\over 2}, {\pi\over 2})$ quasiparticles generate
long--ranged spin distortions similar to those produced by a ferromagnetic
bond, the success of the above--mentioned Aharony model bodes well for
any theory predicting that these carriers exist in the hole--doped
214 systems). If the momentum
of the single--hole ground state is $\vec k = (\pi, 0)$, 
something that may be accomplished through the use of the 
$t - t^\prime - J$ model (see below) with $t^\prime / t > 0$, 
then the tendency towards incommensurability is eliminated and  
only commensurate ordering remains. Clearly, this is 
similar to the above--mentioned behaviour of the hole and electron--doped materials,
and in this paper we will make this comparison complete. 

Very shortly after \cite {GVL}, Tohyama and Maekawa \cite {tohyama1} also studied
the $t - t^\prime - J$ model, now for the electron and hole--doped
systems. They suggested, consistent with a considerable amount of electronic
structure work \cite {hybertsen,eskes,tohyama2}, as well
as angle--resolved photoemission spectroscopy results for the
electron--doped system \cite {allen}, that the {\em Nd} system corresponded to
hopping integrals $t$ and $t^\prime$  with (approximately)
just flipped signs in comparison to the same hopping integrals believed
to be appropriate for the hole--doped
materials (we elaborate on this relation in the next section). 
Then, by examining the energy to the first excited
states, they demonstrated that the commensurate
antiferromagnetic order was much more stable for the electron--doped
system than for the hole--doped systems. As we show below,
this is complimentary to our work in \cite {GVL}, since the single hole
ground state momentum for their electron--doped $t - t^\prime - J$
Hamiltonian is actually $\vec k = (\pi, 0)$! We then strengthen
the arguments of Tohyama and Maekawa by evaluating the magnetic
structure factor $S(\vec q)$ for the many--carrier ground states for
these two systems. Indeed, using the relevant material parameters,
we find incommensurate ordering tendencies for
the hole--doped material, while for the electron--doped material
the magnetic ordering is always commensurate, in complete agreement
with experiment.

We shall also be concerned with other physical properties of the hole
and electron--doped materials. For the $La$ cuprates, some experiments
can be explained beginning with the assumption that at low doping levels
Shraiman--Siggia \cite {SS} dipolar quasiparticles are present (implicit
in these theories is the existence of hole pockets, or a ``small" fermi
surface, something that so far has not been observed). One such magnetic
problem involves the scaling of the correlation length with doping and
temperature, viz. $\xi^{-1} (x,T) = \xi^{-1} (x,0)~+~\xi^{-1} (0,T)$,
and this has been found to be reproduced by a model (based on
the existence of these dipolar quasiparticles) proposed by one 
of us \cite {nskyrmions} for both mobile and localized carriers,
and is in close agreement with experiment \cite {keimer}.
Further, a theory \cite {commentmillis} of the temperature and doping dependent 
magnetic susceptibility  $\chi (x,T)$ has been found to be consistent with 
experiment \cite {johnston}. 
Lastly, the same theory \cite {subir} also predicts the
existence of a pseudogap, something that may be consistent with 
experiment \cite {timusk}. 

We are proposing that these dipolar quasiparticles do not exist in
the weakly--doped $Nd$ cuprates, and thus, e.g., $\xi (x,T),~\chi (x,T)$,
and the optical properties of this system, may be quite different from
those of the $La$ cuprates. Unfortunately, the collection of experiments
performed on the electron--doped materials is not as complete
as the set performed for the hole--doped system, and so
a full comparison is not possible. However, there is one striking difference
between the normal state properties of these systems, and that is the
resistivities at optimal doping. The $La$ system shows a linear--$T$
resistivity \cite {batlogg}, whereas the $Nd$ system shows a clear
fermi--liquid--like  $T^2$ dependence \cite {hikada}. 
Another intriguing experimental result for the electron--doped material
corresponds to the mid--gap
states seen in the angle--integrated photoemission work of Anderson $et~al.$ \cite {allen}.
It has been proposed \cite {emery2} that this is a signal of
phase separation, and to this end we have examined the spatial
distributions of many carriers, as well measured the interaction energies
between carriers. Our results tend to support a phase separation hypothesis, but {\it only} for the
electron--doped system.

Our paper is organized as follows. In \S~\ref{sec:Ham} we discuss the formal aspects of
the $t -t^\prime - J$ model and display the electron--hole symmetries and asymmetries that
can be identified when these materials are doped away from half filling. 
In \S~\ref{sec:numerics} we summarize previously published and our recent 
exact diagonalization results for the one--carrier ground states for these two systems,
and show that these two systems have different one--carrier ground states. 
Then, the spin correlations found in $S(\vec q)$ as 
a function of doping are shown to be in agreement with the phase diagram of Fig. 1.
In \S~\ref{sec:analysis} we examine the spin distortions produced
by each of these quasiparticles via our semi--classical
analysis of this Hamiltonian for a single carrier; this will be shown to agree
with the exact diagonalization results.
Quantum fluctuations are shown to not affect these conclusions. Then we examine the
spatial distribution of carriers pointing out the acute difference between the
hole and electron--doped systems --- from this we relate the tendency towards clustering of
carriers of the electron--doped system to the mid--gap states seen in photoemission.
Finally, in \S~\ref{sec:discussion} we discuss the relevance of our results to the
transport properties of both kinds of systems,
and as a prelude to possible future work 
we consider the extrapolation of this
approach to other systems, and in particular consider the commensurability
that persists in the hole--doped $YBaCuO$ system.

\section{$\lowercase{t} -\lowercase{t}^\prime - J$ model:}
\label{sec:Ham}

\subsection{Electron--Hole Asymmetry induced by $\lowercase{t}^\prime$:} 
\label{sect:asymm1D}

We begin our discussion of the $t - t^\prime - J$ model by first considering
a simpler model involving only hopping terms to the nearest and next nearest
neighbour sites along a one--dimensional chain --- this model provides 
the most direct demonstration of
the electron--hole asymmetry about half filling that $t^\prime$ introduces
into the problem.

Consider a linear chain with a single orbital per site, and define the vacuum 
to be the state with all orbitals devoid of electrons. Introduce
\begin{equation}
H_{hop} = -~t\sum_{<i,j>_{nn}~\sigma} \Big( c_{i\sigma}^\dagger 
c_{j\sigma} + h.c. \Big) 
-~t^\prime\sum_{<i,k>_{nnn}~\sigma} \Big( c_{i\sigma}^\dagger 
c_{k\sigma} + h.c. \Big) 
\label{equation:hopping}
\end{equation}
where $<~>_{nn}$ denotes nearest neighbours, $<~>_{nnn}$ next nearest
neighbours, and $c_{i\sigma}$ is the electron annihilation operator
for site $i$ and spin $\sigma$ (note that these operators are {\it not}
constrained to prohibit double occupancy --- see \S~\ref{sect:sch}). 
Then imagine that the system
has an up spin at every site except one, at which
no electron is present. The energy eigenvalues for 
this hole plus spin polarized background are given by
\begin{equation}
E (k) = - 2 t \cos (k) + 2 t^\prime \cos (2k)~~~~~~~({\rm hole}).
\label{equation:holettp}
\end{equation}
Now consider a system with up spins at every site,
and that in addition to these electrons one down--spin electron is placed
in some orbital. Then the energy eigenvalues of this composite spin--polarized
plus mobile electron system are
\begin{equation}
E (k) = - 2 t \cos (k) - 2 t^\prime \cos (2k)~~~~~~~({\rm electron}).
\label{equation:electronttp}
\end{equation}

From these results it
is to be noted that while the near--neighbour hopping leads to identical
energies for both the hole and electron--doped systems,
the next--nearest--neighbour hopping has different signs.
This reflects the breaking of electron--hole symmetry that $t^\prime$
induces, and thus its inclusion in our starting Hamiltonian is crucial.
It is to be stressed that this asymmetry is found in any dimension, for
any spin background, but only for doping levels close to half filling.

\subsection{Strong--Coupling Hamiltonian:}
\label{sect:sch}

Equation ~(\ref{equation:hopping}) is not a good starting point
for describing the physics of the $CuO_2$ planes in either
$LaSrCuO$ or $NdCeCuO$, since it only describes free carriers.
If instead one begins with an extended Hubbard
model description \cite {emery1} of the  strong correlation
effects known to be present in these systems, and then
examines the systems at close to half filling, one can obtain
the so--called ${t} -{t}^\prime - J$ model \cite {dagotto}. 
To be specific, one considers
\begin{equation}
H = 
-~t\sum_{<i,j>_{nn}~\sigma} \Big( \tilde c_{i\sigma}^\dagger 
\tilde c_{j\sigma} + h.c. \Big) 
-~t^\prime\sum_{<i,k>_{nnn}~\sigma} \Big( \tilde c_{i\sigma}^\dagger 
\tilde c_{k\sigma} + h.c. \Big) 
+ J \sum_{<i,j>_{nn}} 
(\vec S_i \cdot \vec S_j - {1\over 4} n_i n_j)
\label{equation:ttpJ}
\end{equation}
where $J$ is the Heisenberg superexchange constant 
coupling spins $\vec S_i$ between nearest--neighbour sites.
Also, in Eq.~(\ref{equation:ttpJ}) constrained electron operators are 
used (constrained to disallow double occupancy at any site),
viz.
\begin{equation}
\tilde c_{i\sigma} = c_{i\sigma} ( 1 - c^\dagger_{i-\sigma}c_{i-\sigma} )~~.
\label{equation:contrainedcs}
\end{equation}

We wish to stress that our use of the ${t} - {t}^\prime - J$ model is based
on beginning with a three--band extended Hubbard model description \cite {emery1} 
for {\em both} the $LaSrCuO$ and $NdCeCuO$ systems. While it is well 
known \cite {zhangrice,shastry,goodingelser} 
that for the hole--doped systems one may approximately
map the three-band Hubbard model onto the one--band
$t - t^\prime - J$ model (N.B. --- not necessarily the one--band Hubbard model), 
it may be argued that this is
also the correct starting point for the electron--doped systems. For example,
the dominant contribution to the $t^\prime$ hopping integral for the
electron doped system is {\em not} due to a direct $Cu - Cu$ overlap, but rather follows
from a third--order $Cu \rightarrow O \rightarrow O \rightarrow Cu$ hopping
\cite {schlueter}. Thus, in what follows we consider (i) the $t$ and $t^\prime$
parameters that are to be {\em fitted} via quantum cluster studies (see below),
and not due to a strong--coupling mapping of the one--band Hubbard model
onto a Hibert space prohibiting double occupancy \cite {trugman}, and (ii)
the superexchange constant $J$ to be found from either quantum cluster
studies, or from comparison to experiments probing the magnetic
structure and/or excitations.

In the previous subsection we were able to treat hole and electron--doped
systems with the same starting Hamiltonian. However, now
it is clear that the above Hamiltonian, for a vacuum corresponding to no electrons at
any site, cannot be used to describe both hole and electron doping of the
half--filled state. For example, a vacancy can easily be added to a spin--polarized
ferromagnetic state, and is completely mobile, while a down spin electron
added to such a spin background is completely localized. Consequently, with
this Hamiltonian and this vacuum, only hole--doped $CuO_2$ planes may be
investigated.

To investigate the electron--doped system
it is usual to introduce an electron--hole
transformation, viz.
\begin{equation}
c_{i\sigma} \rightarrow a^\dagger_{i\sigma}
\label{equation:eh1}
\end{equation}
where $a_{i\sigma}$ is the annihilation operator of a hole at site $i$ of 
spin $\sigma$. (With this transformation it follows that one now
considers a vacuum with no holes at any site.)
However, this does not lead to the desired Hamiltonian
as one finds that the new hopping term does not describe constrained (viz. two holes
are never allowed on the same site) hopping. Instead, one must utilize
a more general model than Eq.~(\ref{equation:ttpJ}), one that includes
so--called doublon hopping processes \cite {clarke}, and from this
starting Hamiltonian it is straightforward to derive the required
transformation.
The result of this transformation can then be shown to be equivalent to
\begin{equation}
\tilde c_{i\sigma} \rightarrow \tilde a^\dagger_{i\sigma}
\label{equation:eh2}
\end{equation}
where the $\tilde a_{i\sigma}$ also satisfy fermionic anti--commutation relations.
Thus,  one eventually finds that the appropriate $t - t^\prime - J$ model for the 
electron--doped system is the same as Eq.~(\ref{equation:ttpJ}) but with
the $\tilde a_{i\sigma}$ replacing the $\tilde c_{i\sigma}$ {\it and} with
the minus signs in front of $t$ and $t^\prime$ changed to be plus signs.

Summarizing the above discussion, the $t - t^\prime - J$ Hamiltonian will be used 
to model the physical systems that we are interested in. 
We consider this model to possess the bare minimum number of processes
through which both the hole and electron--doped $CuO_2$ planes can be
represented. To be specific, 
(i) Eq. ~(\ref{equation:ttpJ}) with a vacuum
of zero electrons at every site of a square lattice will describe the hole--doped
system, and  (ii) for the electron--doped systems we use
\begin{equation}
H = 
~t\sum_{<i,j>_{nn}~\sigma} \Big( \tilde a_{i\sigma}^\dagger 
\tilde a_{j\sigma} + h.c. \Big) 
+~t^\prime\sum_{<i,k>_{nnn}~\sigma} \Big( \tilde a_{i\sigma}^\dagger 
\tilde a_{k\sigma} + h.c. \Big) 
+ J \sum_{<i,j>_{nn}} 
(\vec S_i \cdot \vec S_j - {1\over 4} n_i n_j)
\label{equation:ttpJel}
\end{equation}
For this latter Hamiltonian we use a vacuum of zero holes at every 
site of a square lattice.  Thus, note that, quite simply, in comparison to 
the hole--doped Hamiltonian the signs of the $t$ and $t^\prime$ hopping 
terms are flipped and the $\tilde a$ operators replace
the $\tilde c$ operators. Thus, from now on we simply use 
Eq.~(\ref{equation:ttpJ}) for both systems, specifying $t=1$ for 
hole doping, and $t=-1$ for electron doping \cite {tohyama1}.

We now consider the numerical values of the hopping and superexchange
parameters.
A consequence of the holes residing on the $O$ sites in the $La$ 
material is that for a 
quantitative determination of the hopping parameters $t$ and $t^\prime$, 
something very different from overlap integrals must be evaluated.
We refer the interested reader to the 
summary of this problem given by Tohyama and Maekawa \cite{tohyama1},
and simply state the range of accepted values \cite {tohyama1,hybertsen,eskes}.
For the hole--doped $LaSrCuO$ system, using Eq.~(\ref{equation:ttpJ}),
one may scale all energies such that $t=1$. Then one has 
$J \approx .3 \rightarrow .4$
and $t^\prime \approx -.2 \rightarrow -.4$. For the electron--doped 
$NdCeCuO$ system we again use Eq.~(\ref{equation:ttpJ}) but now with $t= - 1$,
and $J \approx .3 \rightarrow .4$ and $t^\prime \approx + .2 \rightarrow + .4$.
Thus, it is interesting to note that even 
though a Zhang--Rice singlet is not exactly equivalent
to a $Cu~3d^8$ ion, the material parameters for these two systems are
such that the electron--hole mapping given in Eq.~(\ref{equation:eh2})
predicts reasonably well the  numerical values that one must employ
for one system given the parameters of the other.

\section{Numerical Results:}
\label{sec:numerics}

Stephan and Horsch \cite {stephan} have proposed that
knowledge of the single--hole system does not lead to valuable
information on the multiply--doped state. In \cite {GVL} we
showed that this is not always the case, and that, in particular,
knowledge of the one--hole ground state can aid in predicting the
presence/absence of incommensurate correlations. Thus, we begin
a discussion of our numerical work with a summary of the single--hole
ground state.

There has already been a comprehensive exact diagonalization study 
of the ground state of one hole in the $t - t^\prime - J$ model for 
a $4 \times 4$ square cluster in Ref. \cite {gagliano} (since they use a bosonic
representation, their sign of $t^\prime$ is flipped in comparison
to ours). For the physically relevant ratio of $J/t$ they find a $\vec k
= \pm ({\pi\over 2}, \pm {\pi\over 2})$ ground state when
the $t^\prime$ corresponding to hole--doped materials is used (unless 
$\mid t^\prime \mid$ becomes too large), and a 
$\vec k = \pm (\pi,0), \pm (0,\pi)$ ground state
for the $t^\prime$ associated with the electron--doped materials. 

We have used a Lanczos routine and have also performed exact diagonalization
evaluations \cite {leung} of the ground states of the $t - t^\prime - J$ model
for $J=.4$, with $t = \pm 1$ for $t^\prime = \pm .1, \pm .2$ and $ \pm .3$.
For the conventional $4 \times 4$ 16--site square cluster our
work agrees with that of Ref. \cite {gagliano}. We have also
studied the $\sqrt {8} \times \sqrt {32}$ 16--site cluster introduced
in \cite {GVL}, as well as some results generated using
a $\sqrt {18} \times \sqrt {32}$ 24--site cluster \cite {finitesize}.
For the hole--doped material, viz. when $t^\prime < 0$, we always 
find a ground state of $\vec k = \pm ({\pi\over 2}, \pm {\pi\over 2})$,
while for the electron--doped material, viz. when $t^\prime > 0$, we always \cite {24site}
find a ground state of $\vec k = (\pm\pi, 0),~(0,\pm\pi)$. Our
results for these two different clusters are entirely consistent 
with the phase diagram of Ref. \cite {gagliano} found using the $4 \times 4$
cluster.

Our numerical results also lead us to expect that the band structures
for the hole and electron--doped materials
will be quite different. We find that the band structure of the hole--doped material
is similar to Fig. 4 of \cite {GVL} (which is a $t^\prime = 0$ curve), 
although with the increased ratio of $\mid t^\prime  / t \mid$ that we are
using here the band along $\vec k = (0,0) \rightarrow (\pi,0)$
becomes quite flat.  This is similar to recent photoemission work of
the valence band of undoped $Sr_2CuO_2Cl_2$ \cite {wells}, and provides 
some experimental support for a value of $t^\prime$ around $-0.3~t$. 
We have listed the single carrier band structure energies for our 
${\sqrt 18} \times {\sqrt 32}$ 24--site cluster in Table I for both the hole 
and electron--doped systems having used $t^\prime / t~=~-0.3.$

The single--hole ground states of the hole and electron--doped
materials are different --- so what? As mentioned
above, the important fact contained in Ref. \cite {GVL} is that
a knowledge of the single--hole
ground state can aid in understanding whether or not any
short--ranged incommensurate correlations are introduced into
the ground state of the multiply--doped system. This
followed from (i) if the single--hole ground state is not
a $\vec k = \pm ({\pi\over 2}, \pm {\pi\over 2})$ state,
no tendencies towards incommensurate correlations occurred,
consistent with the spiral phase prediction of
Shraiman and Siggia \cite {spiral},
and (ii) the electron momentum distribution function for
2, 3 and 4 holes was found to be composed of 
the half--filled fermi surface with dimples at those momenta corresponding
to the single--hole ground states, thus suggesting that
some form of rigid band filling \cite {GVL} is in effect.  Combining
this with the differing single--hole ground states mentioned
above, it is clear that we can now extrapolate these abstract results
to the physical systems under consideration.

Figures 2 and 3 show the magnetic structure factor for 0, 1, 2, 3, and 4
holes (electrons) for $t^\prime = \pm .3$ evaluated with the 
$\sqrt {8} \times \sqrt {32}$ cluster introduced in \cite {GVL} ---
similar results are obtained for other $\pm t^\prime$ pairs.
For the hole--doped material, Fig. 2 displays that the peak in the 
structure factor shifts from $(\pi,\pi)$ for 0 and 1 hole, to 
$({3\pi\over 4},{3\pi\over 4})$ for
2 holes, to $({\pi\over 2},{\pi\over 2})$ for 3 and 4 holes. This behaviour
is similar to the behaviour displayed in Fig. 5 of Ref. \cite {GVL}
for $t^\prime = 0$. However, in Fig. 3 it is seen that no
shifts of the maximum of $S(\vec q)$ away from $(\pi,\pi)$ occurs for
any doping concentration away from half filling for the electron--doped system.
These results are entirely consistent with Fig. 1,
and, in particular, are consistent with the lack of any tendency of the electron--doped
materials to display the kind of incommensurate correlations
that are present in the hole--doped system \cite {incomdirection}.
Clearly, these numerical results suggest that we have found
a reasonable starting point from which one can hope to be able to 
describe the hole and electron--doped 214 systems. We now try to
come to an understanding of these numerical results.

\section{Analysis:}
\label{sec:analysis}

As mentioned in the introduction, the existence of Shraiman--Siggia
quasiparticles in weakly doped $LaSrCuO$ is supported
by the theories \cite {polarons,nskyrmions,commentmillis,subir,noha} 
that assume their existence, and are subsequently
able to produce results consistent with experiment. To be specific,
it seems that a scenario in which these quasiparticles are
weakly interacting and tend to stay very far apart from one another
(thus making it likely that some form of rigid band filling (e.g.
with hole pockets around $\pm({\pi\over 2},\pm{\pi\over 2})$) is applicable. 

Now we study the $(\pi,0)$ quasiparticles: we are proposing that
these are the single--hole ground state constituents that
exist in the weakly doped $NdCeCuO$ system. We wish to characterize
the spin distortions that each such quasiparticle produces,
as well as come to an understanding of the spatial distribution
for a state which is multiply doped. This will allow us to (i)
understand why the $(\pi,0)$ and $({\pi\over 2}, {\pi\over 2})$
quasiparticles lead to differing
magnetic structure factors vs. doping, and (ii) examine
some of the other physical features that are known to be specific
to the electron--doped system.

According to the semiclassical theory of Shraiman and Siggia \cite {SS},
for ground states with these wave vectors, differing distortions
of the spin background occur. For the $\vec k = ({\pi\over 2},
{\pi\over 2})$ states, long--ranged spin distortions with
dipolar symmetry are produced via backflow.
For the $\vec k = (\pi,0)$ ground states, the hydrodynamic
theory of Shraiman and Siggia \cite {SS} leads to the prediction that 
the spin distortion induced by one $(\pi,0)$ quasiparticle
is short ranged. These ideas are verified in our Figs. 4 and 5
where our results from evaluating
the energies at this wave vector for the $t - t^\prime - J$ model
with the electron--doped material parameters using a semiclassical
variational wave function, such as that described in \cite {SS},
are displayed.
To be specific, we utilize a product state of classical spins
in an infinite lattice, and incorporate a broken antiferromagnetic
symmetry in calculations of the hopping matrix elements for
a single hole (see \cite {SS} for the details of such a
variational wave function). Then, to display the range of
the spin distortions for this ground state we have calculated
the minimum energy as a function of the number of spins away
from the hole that are allowed to be distorted from their
undoped Neel configuration --- these sites are shown in Fig. 4, where
they are labeled according to the
equivalent sites around a single carrier. 
In Fig. 5 we show the minimum energies found from the
variational principle
as a function of these site labels for both the hole--doped
$\vec k = ({\pi\over 2},{\pi\over 2})$ one hole ground state, and
the $\vec k = (\pi,0)$ electron--doped one carrier ground state. 
The long--ranged spin distortion of the hole--doped ground
state is clear, whereas the short--ranged distortion of the
electron--doped ground state is made manifest by the
insensitivity of the minimum energy when one allows more
distant neighbours to be distorted from their Neel state.

The above procedure may be repeated for all wave vectors
for both hole and electron--doped Hamiltonians. We find
that providing $t^\prime$ satisfies certain inequalities
(which necessarily depend on the ratio of $J/t$) that the
semiclassical band structure largely reproduces the
band structures determined by exact diagonalization. Specifically,
the ground states are reproduced when (i) $t^\prime > -.12$
for the hole--doped material, and (ii) $t^\prime > +.12$
for the electron--doped materials (these inequalities
will be affected by quantum fluctuations --- see below). 
This agreement with the quantum cluster studies provides further 
evidence that the renormalized classical description of the spin 
background ({\it not} including the charge carriers) of the weakly doped systems is valid at least
at low temperatures \cite {birgeneau}.

This semiclassical work ignores quantum 
fluctuations.  However, as shown by Reiter \cite {reiter}, 
for an infinite lattice when quantum fluctuations are included in such
variational wave functions one still finds the Shraiman--Siggia
dipolar quasiparticles with the long--ranged dipolar spin
distortion intact. We have found \cite {unpublished1} that within
this lowest--order self--consistent approximation \cite {reiter} 
the $(\pi,0)$ states still only involve short--ranged spin distortions,
the same as the semiclassical prediction. Thus, it
is not unexpected that our quantum cluster and semiclassical analyses agree. 

Now let us consider the effect of the short--ranged spin distortions 
that this kind of quasiparticle produces, viz. why do these
quasiparticles not introduce any incommensurate correlations into the
spin texture. Given the short--ranged nature of the spin distortion,
it is clear that they could lead 
to a disturbance of the spin texture similar to that found in 
the quantum dilution problem \cite {manousakis}, since static vacancies 
also produce a short--ranged disturbance of the spin texture.
However, the quantum dilution problem begins with the assumption
that the dilutants are randomly distributed throughout a plane.
We now show that this is probably not the case for the mobile carriers
in the electron--doped materials.

In Ref. \cite {GVL} we presented some numerics for the hole--hole 
correlation function for a pair of holes doped into a 
$\sqrt {8} \times \sqrt {32}$ cluster described by the $t - J$ model.
We found that the holes tended to stay as far apart as possible.
(It is interesting to note that this is entirely consistent 
with experimental work purporting to
see charge--rich walls, separated by distances scaling like $1/x$,
arising from finite--size striped magnetic domains at low $x$ \cite {cho}.
Thus, our work demonstrates a magnetic mechanism for such stripes,
and is different than the Coulomb--interaction generated carrier--rich
walls of Emery and Kivelson's \cite {emery2} phase separation ideas.)
Since this numerical result is in direct contradiction to the results
of Poilblanc \cite {poilblanc2}, who studied pairs of holes in a variety of square
clusters, we have further scrutinized this behaviour. We have found
that in (i) Monte Carlo simulations of the $t - U$ model, for an
average of 2 holes in an $8 \times 8$ lattice, the holes still
tend to want to be as far apart as possible, and (ii) the same
behaviour is found for the larger $\sqrt {18} \times \sqrt {32}$
24--site cluster with 2 holes --- these results will be published in
a future comment \cite {chen}. 

We now present our results for similar calculations for the 
electron--doped system. Figure 6 shows the carrier--carrier 
correlation function, defined analogously to the hole--hole correlation
function of Eq. (6.1) of \cite {GVL}, by
\begin{equation}
P_{cc} (| i - j|) = {1\over N_T N_e} \sum_{i,j} <
(1 - \sum_{\sigma} n_{i, \sigma} ) (1 - \sum_{\sigma} n_{j, \sigma} ) >
\end{equation}
where $n_{i, \sigma}$ is the hole number operator for electron--doped
systems (electron number operator for hole--doped
systems, as in Eq. (6.1) of \cite {GVL}) for spin $\sigma$ at site $i$,
$N_T$ being the number of lattice sites,
and $N_e$ is the number of equivalent sites a distance $|i - j|$ from site $i$
(note that we do not include the angular dependence which is actually present
for our non-square lattice, for simplicity only, since we have found that
this does not affect our conclusions). We evaluated this function
for two carriers doped into a $\sqrt {8} \times \sqrt {32}$ cluster
described by the $t - t^\prime - J$ model with $t=-1,~J=.4~,$ and $t^\prime = +.3$.
Juxtaposed with this curve is the analogous hole-hole correlation function
for $t=1,~J=.4,$ and $t^\prime = -.3$. The differences between the two
different systems is striking: whereas pairs of holes tend to stay as far
apart as possible, pairs of mobile electrons in the electron--doped system tend
to cluster together.

To further elucidate this behaviour we have
calculated the 2, 3 and 4 carrier interaction energies (analogous to
the binding energy) defined relative to {\it independent} carriers,
(denoted by $E_{2I},~E_{3I}$ and $E_{4I}$, where $E_n$ denotes the ground
state for $n$ carriers), via
\begin{eqnarray}
E_2 &= E_0 + 2 (E_1 - E_0) + E_{2I} \nonumber \\
E_3 &= E_0 + 3 (E_1 - E_0) + E_{3I} \nonumber \\
E_4 &= E_0 + 4 (E_1 - E_0) + E_{4I} 
\label{equation:binding}
\end{eqnarray}
and our results for both hole and electron--doped systems are presented in Table II.
One sees that for the hole--doped system, the tendency for holes to cluster
together is very small --- this is consistent with the binding energy study
of Riera and Young \cite {riera} (if one uses their definition for the binding
energy one finds results very similar to those of Table II). However, for
the electron--doped material one finds that there is both a stronger binding
of a single pair of holes, as well as a very low interaction energy, at least relative
to the hole--doped material, as the number of carriers increases. This shows
that the mobile electrons have a much greater propensity to cluster together
than do the mobile holes.

Summarizing the numerical work of the last two sections: (i) The
single--hole quasiparticles of the hole--doped system are Shraiman--Siggia
dipolar quasiparticles. They tend to remain far from each other,
and are thus consistent with some form of superposition of these many--body
quantum states in the low--doping limit \cite {GVL}. (ii) The single--carrier
ground state 
of the electron--doped material is different than that of the hole--doped system,
viz. it is a $(\pi,0)$ state as opposed to a $({\pi\over 2}, {\pi\over 2})$ state,
and a non--zero density of the former tend to 
exhibit a much stronger tendency to cluster together than do
the holes of the hole--doped material. The spin distortions for these
quasiparticles are long (short) ranged for the hole (electron) doped systems.

Considering the data contained in Fig. 6 and Table II, we have reason to believe
that for the electron--doped materials the carriers have a strong tendency
towards phase separation --- certainly the tendency is much stronger
than in the hole--doped system. However, since Table II shows that 
there are not purely attractive
interactions between the mobile electron carriers, this tendency will probably
lead to the presence of low--energy excited states corresponding to
fluctuations into a locally phase separated state. Then, perhaps these
phase separated states are the cause of the mid--gap states seen in the
photoemission work \cite {allen}, as was suggested elsewhere \cite {emery2}.

As espoused by Emery and Kivelson \cite {emery2}, phase separation in the
large hopping limit is likely to be frustrated, the frustration being due to
the presence of long--ranged Coulomb interactions. We have investigated the
effect of including a near--neighbour Coulomb repulsion between carriers
on neighbouring sites, and find that no qualitative changes take place until
very large Coulomb interactions (say of order much larger than $t$) are
present. However, we do not know how long--ranged Coulomb interactions
will effect our conclusions, and this is presently being studied.
Further, Trugman \cite {trugman} noted that tendencies towards the binding of a pair
of holes can be greatly reduced by next--nearest neighbour three--site 
spin--dependent hopping.
While this form of hopping will be of much lower order than the next--nearest
neighbour hopping that we are studying (since we begin with a three-band,
not a one--band (as does Trugman), description), and thus we do not expect
any qualitative changes in our analysis, we feel it appropriate
to see in what quantitative ways our observed tendency towards a phase separated state in 
electron--doped planes changes with such a hopping term present. Thus, we are
also presently studying the effects of such terms.

Consequently, we propose
that the commensurate antiferromagnetic spin background of the electron--doped materials do
not have incommensurate correlations introduced into them because (i) the quasiparticles
introduced into the weakly doped planes are not Shraiman--Siggia dipolar quasiparticles,
and (ii) the mobile electrons found in these materials exhibit a form of
phase separation, and no incommensurate correlations are expected for
such a state of matter. We have completed a study of the electron/hole momentum
distribution functions, similar to those presented in Ref. \cite {GVL}, including an
evaluation of the density of states for both the hole and electron--doped
compounds, and this work (to be presented later)
further strengthens our assertion that some form of carrier clustering is
taking place in the electron--doped materials.

\section{Discussion:}
\label{sec:discussion}

Our numerical results and semiclassical studies have led to 
the proposal that the single--hole ground
states in the hole and electron--doped materials are different, and that 
this has a profound effect on (i) their magnetic phase diagrams, and (ii) the tendency
towards phase separation that these carriers display. The Shraiman--Siggia
dipolar quasiparticle picture has been shown \cite {nskyrmions,commentmillis,subir}
to be consistent with many different experiments studying the hole--doped
$LaSrCuO$ system, and it will be interesting to see if our $(\pi,0)$ quasiparticle
picture of phase separation in the $NdCeCuO$ can be used to
explain more than just the ARPES results associated with 
the mid--gap states found in ARPES studies \cite {allen}. 

One other experiment that may possibly find an explanation in the frustrated
phase separation scenario for the electron--doped material
involves the resistivity vs. temperature results
of Hikada $et~al.$ \cite {hikada}. To be specific, if the mobile electrons
tend to cluster together for sufficiently long periods of time, the dominant
interaction between carriers is electronic as opposed to magnetic,
and thus a fermi--liquid--like $\rho~\sim~T^2$ is not that surprising.
Of course, such a qualitative explanation requires rigorous
calculations to be carried out before it is to be taken seriously.

We have addressed the magnetic and other material characteristics
of the two 214 high $T_c$ systems. However, the approach of utilizing
the $t - t^\prime - J$ system does not seem to be limited to these 
single--layer high $T_c$ materials. Recently, it has been proposed
\cite {chub123} that hole pockets appear in the $YBaCuO$ material
as one dopes away from the antiferromagnetic insulator, and that
these pockets are centred at $({\pi\over 2},{\pi\over 2})$. This is entirely consistent
with the ``small fermi surface" observed in ARPES studies
of $YBa_2Cu_3O_{6.3}$ \cite {liu}.
In order to model such a system one must include a $t^\prime$
parameter that rotates the fermi surface by 45$^\circ$ with respect
to that predicted by the $t - J$ model, and to this end we
have used $t=1,~t^\prime\sim-.5,$ and $J=.4$. Then, we find (i)
that indeed the single--hole ground state occurs at $\vec k = (\pi,\pi)$
(this was also found in the study of Ref. \cite {gagliano}),
and (ii) that pairs of such holes tend to stay as far apart as
possible, with a hole--hole correlation function similar to that
shown for the hole--doped material in Fig. 6. Thus, weakly interacting
quasiparticles forming small hole pockets at $(\pi,\pi)$ 
is consistent with our numerics, and are not expected to lead
to any incommensurability \cite {chub123}. Noting that both our
semiclassical work (similar to that shown in Fig. 5) and the spin--wave
analysis of Reiter \cite {reiter} predict that the spin distortions produced
by $(\pi,\pi)$ quasiparticles are long ranged, it is very interesting
that in the two systems in which the holes tend to stay as far
apart as possible, the single--carrier quasiparticles produce long--ranged
spin distortions, and we are presently exploring this coincidence.

\bigskip
\bigskip
\bigskip
\bigskip
\bigskip
\centerline{\bf {ACKNOWLEDGEMENTS:}}

We are very grateful to Elbio Dagotto for providing us a copy of Fig. 1,
and for sending us a copy of Ref. \cite {dagotto} prior
to publication.  We also wish to thank Andrey Chubukov for discussions
concerning the 123 material, and Liang Chen for helpful
comments on the hole--hole correlation function for the $t -U$ model. 
Lastly, we wish to thank Vic Emery and Barry Wells for conveying to
us important information on the $Sr_2CuO_2Cl_2$ photoemission data.
This work was supported by the NSERC of Canada.

\par\vfill\eject

\begin{figure}
\caption{A schematic of the phase diagrams, as a function of temperature and
doping concentrations away from half filling, for both the hole 
and electron--doped materials (after Ref. [36], from Ref. [4]).}
\label{fig:phasediagram}
\end{figure}

\begin{figure}
\caption{The magnetic structure factor
for the $t - t^\prime - J$ model given in Eq.~(2.4) for 0, 
1, 2, 3 and 4 carriers; here these carriers are mobile holes. 
For this figure, $t = +1,~t^\prime = -.3,$ and $J = .4$, which are appropriate 
for hole--doped $LaSrCuO$.  The reciprocal lattice points are as follows: 
$\Gamma=(0,0),~X=(\pi,0),$ and M=$(\pi,\pi)$.}
\label{fig:sofqholes}
\end{figure}

\begin{figure}
\caption{The magnetic structure factor
for the $t - t^\prime -J$ model given in Eq.~(2.4) for 0, 1, 2, 3 
and 4 carriers; here these carriers are mobile electrons.
For this figure, $t = -1,~t^\prime = +.3,$ and $J = .4$, which are appropriate
for electron--doped $NdCeCuO$.}
\label{fig:sofqelectrons}
\end{figure}

\begin{figure}
\caption{A picture of the equivalent sites that we allow to be distorted away
from their ordered N\'eel configuration in our calculation of the semiclassical
wave function for the $t - t^\prime - J$ model for one carrier. The clusters used in
such calculations are infinite in extent.}
\label{fig:semicllattice}
\end{figure}

\begin{figure}
\caption{The energies of the $\vec k = (\pi,0)$ ground state for the electron--doped 
system ($t = -1,~t^\prime = +.3,~J = .4$) and the
$\vec k = ({\pi\over 2}, {\pi\over 2})$ ground state for the hole--doped 
system ($t = +1,~t^\prime = -.3,~J = .4$) obtained via application of
the variational principle to a Shraiman--Siggia semiclassical wave function.
These energies are found by allowing all spins up to distance $d$, given
in units of the lattice constant $a$, to be distorted away from their
antiferromagnetically aligned state.}
\label{fig:semiclenergy}
\end{figure}

\begin{figure}
\caption{The carrier--carrier correlation function, defined and described in Eq.~(6.1) of
Ref. [18],  for the hole ($t^\prime = -.3$) and electron ($t^\prime = +.3$) doped
systems, evaluated for the doubly doped state. It is seen that the holes tend to stay
as far apart as possible, whereas the mobile electrons tend to cluster together
at short distances.}
\label{fig:holehole}
\end{figure}

\par\vfill\eject
\begin{table}
\caption{Single carrier minimum energies for both hole and electron singly--doped
planes, described by Eq.~(2.4), for the allowed wave vectors of our non--square
24--site cluster. All energies are in units of $t$.}
\begin{tabular}{lcc}
~~~~$\vec k$&Energy &Energy\\
&(Hole doped: $t = 1 {,}~t^\prime = -.3$)&(Electron doped: $t = -1 {,}~t^\prime = +.3$)\\
\tableline
$(0,0)$&-12.824&-12.153\\
$({\pi\over 4},{\pi\over 4})$&-12.961&-12.338\\
$({\pi\over 2},{\pi\over 2})$&-13.207&-12.686\\
$({3\pi\over 4},{3\pi\over 4})$&-13.063&-12.375\\
$(\pi,\pi)$&-12.853&-12.177\\
$({-\pi\over 3},{\pi\over 3})$&-12.997&-12.471\\
$({-2\pi\over 3},{2\pi\over 3})$&-13.127&-12.363\\
$({-\pi\over 12},{7\pi\over 12})$&-12.896&-12.746\\
$({-5\pi\over 12},{11\pi\over 12})$&-12.980&-12.679\\
$({\pi\over 6},{5\pi\over 6})$&-12.821&-13.037\\
\end{tabular}
\end{table}
\vskip 6.0 truecm
\begin{table}
\caption{Interaction energies, defined in Eq.~(4.1), for 2, 3 and 4 carriers
in the $t - t^\prime -J$ model, using representative material parameters
for the hole and electron--doped compounds based on Eq.~(2.4). As 
long as the single--hole ground states
are those that we predict for the hole and electron--doped systems,
these interaction energies do not sensitively
depend on the ratio of $\mid t^\prime / t \mid$.}
\begin{tabular}{lcc}
&Hole Doped&Electron Doped\\
&$t = 1 {,}~t^\prime = -.3$&$t = -1 {,}~t^\prime = +.3$\\
\tableline
E$_{2I}$&-.12&-.21\\
E$_{3I}$&+.45&+.12\\
E$_{4I}$&+1.5&+.26\\
\end{tabular}
\end{table}

\begin{references}
\bibitem{bednorz}  J.G. Bednorz and K.A. M\"uller, Z. Phys. B {\bf 64}, 189 (1986).
\bibitem{tokura} Y. Tokura, H. Takagi, and S. Uchida, Nature {\bf 337}, 345 (1989).
\bibitem{birgeneau} See, e.g., G. Shirane, R.J. Birgeneau, Y. Endoh,
and M.A. Kastner, Physica B {\bf 197}, 158 (1994).
\bibitem{electronmag} See, e.g., C. Almasan and M.B. Maple, in ``Chemistry of
High Temperature Superconductors", ed. by C.N.R. Rao (World Scientific, Singapore, 1992).
\bibitem{aeppli} S.--W. Cheong, G.A. Aeppli, T.E. Mason, H. Mook, S.M. Hayden,
P.C. Canfield, Z. Fisk, K.N. Clausen, and J.L. Martinez, Phys. Rev. Lett. {\bf 67}, 1791 (1991).
\bibitem{mason} T.E. Mason, G. Aeppli, and H.A. Mook, Phys. Rev. Lett. {\bf 68}, 1414 (1992).
\bibitem{thurston2} T.R. Thurston, P.M. Gehring, G. Shirane, R.J. Birgeneau,
M.A. Kastner, Y. Endoh, M. Matsuda, K. Yamada, H. Kojima, and I. Tanaka,
Phys. Rev. B {\bf 46}, 9128 (1992).
\bibitem{aharony} A. Aharony, R.J. Birgeneau, A. Coniglio, M.A. Kastner,
and H.E. Stanley, Phys. Rev. Lett. {\bf 60}, 1330 (1988).
\bibitem{polarons} R.J. Gooding, and A. Mailhot, Phys. Rev. B. {\bf 44}, 11852 (1991).
\bibitem{frenkel} D.M. Frenkel, R.J. Gooding, B.I. Shraiman, and E.D. Siggia, 
Phys. Rev. B. {\bf 41}, 350 (1990).
\bibitem{keimer} B. Keimer, N. Belk, R.J. Birgeneau, A. Cassanho, C.Y. Chen,
M. Greven, M.A. Kastner, A. Aharony, Y. Endoh, R.W. Erwin, and
G. Shirane,  Phys. Rev. B {\bf 46}, 14034 (1992).
\bibitem{manousakis} E. Manousakis, Phys. Rev. B {\bf 45}, 7570 (1992).
\bibitem{thurston1} T.R. Thurston, M. Matsuda, K. Kakurai, Y. Yamada, Y. Endoh, R.J. Birgeneau,
P.M. Gehring, Y. Hikada, M.A. Kastner, T. Murakami, and G. Shirane, Phys. Rev. Lett.
{\bf 65}, 263 (1990).
\bibitem{behre} J. Behre, S. Miyashita and H.J. Mikeska, J. Mag. and Mag. Materials, {\bf 104--107}, 863 (1992).
\bibitem{batlogg} H. Takagi, B. Batlogg, H.L Kao, J. Kwo, R.J. Cava, J.J. Krajewski,
and W.F. Peck, Phys. Rev. Lett. {\bf 69}, 2975 (1992).
\bibitem{uji} S. Uji and H. Aoki, Physica C {\bf 199}, 231 (1992).
\bibitem{second} Even if the carriers were localized, there is another reason why
the above analyses cannot be a complete explanation of the magnetic phase diagrams.
The localized defects discussed above do indeed occur at low
temperatures, but for both systems they arise from the electrostatic
attraction of the carriers to the dopants: either a hole at a {\em O$^-$} site or a
an electron at a {\em Cu$^+$} site, to the inhomogeneous charge background produced
by either a {\em Sr$^{2+}$} or {\em Ce$^{4+}$} impurity, the latter being
embedded in the {\em La$^{3+}$} or {\em Nd$^{3+}$} oxygen layer above
a {\em CuO$_2$} plane. The sites at which the carriers could be
localized are degenerate. In particular, there is an equivalence 
of the 4 sites in the immediate region of the impurity on which the
carrier could localize. Why would the carriers localize
on just one such site when any of the other 3 sites are just
as good? Clearly, the carriers can only be localized to at least
these four sites.  Then, the question is: since the holes
and electrons both distribute themselves around {\em Sr} or {\em Ce}
defects, why is the spin distortion produced by these systems
different? Neither the frustrated--bond (\cite {aharony})
or the quantum dilution (\cite {manousakis}) models possess this symmetry. We have
found an explanation of this problem, based on our earlier work on this problem, viz. 
R.J. Gooding, Phys. Rev. Lett. {\bf 66}, 2266 (1991),
and the details of this will be presented elsewhere.
\bibitem{GVL} R.J. Gooding, K.J.E. Vos, and P.W. Leung, Phys Rev. B {\bf 49}, 4119 (1994).
\bibitem{moreo1} A. Moreo, E. Dagotto, T. Joliceur, and J. Riera, Phys. Rev. B 
{\bf 42}, 6283 (1990).
\bibitem{moreo2} A. Moreo, D.J. Scalapino, R. Sugar, S. White, and N. Bickers, 
Phys. Rev. B {\bf 41}, 2313 (1990).
\bibitem{furukawa} N. Furukawa and M. Imada, J. Phys. Soc. Jpn. {\bf 61}, 3331 (1992).
\bibitem{tohyama1} T. Tohyama and S. Maekawa, Phys. Rev. B {\bf 49}, 3596 (1994).
\bibitem{hybertsen} M.S. Hybertsen, E.B. Stechel, M. Schl\"uter,
and D.R. Jennison, Phys. Rev. B {\bf 41}, 11068 (1990).
\bibitem{eskes} H. Eskes, G.A. Sawatzky, and L.A. Feiner, Physica  C {\bf 160}, 424 (1989).
\bibitem{tohyama2} T. Tohyama and S. Maekawa, J. Phys. Soc. Jpn., {\bf 59}, 1760 (1990).
\bibitem{allen} R.O. Anderson, R. Claessen, J.W. Allen, C.G. Olson, C. Janowitz,
L.Z. Liu, J.--H. Park, M.B. Maple, Y. Dalichaouch, M.C. de Andrade, R.F. Jardim,
E.A. Early, S.--J. Oh, and W.P. Ellis, Phys. Rev. Lett. {\bf 70}, 3163 (1993).
\bibitem{SS} B.I. Shraiman and E.D. Siggia, Phys. Rev. Lett. {\bf 61}, 467 (1988).
\bibitem{nskyrmions} R.J. Gooding and A. Mailhot, Phys. Rev. B {\bf 48}, 6132 (1993).
\bibitem{commentmillis} A.V. Chubukov and S. Sachdev, Phys. Rev. Lett. {\bf 71}, 3615(C) (1993).
\bibitem{johnston} D.C. Johnston, Phys. Rev. Lett. {\bf 62}, 957 (1989).
\bibitem{subir} S. Sachdev, Phys. Rev. B {\bf 49}, 6770 (1994).
\bibitem{timusk} C.C. Homes, T. Timusk, R. Liang, D.A. Bonn, and W.N. Hardy,
Phys. Rev. Lett. {\bf 71}, 1645 (1993); at present, this experiment has not
been performed on an underdoped $La$ cuprate crystal.
\bibitem{hikada} Y. Hikada and M. Suzuki, Nature (London) {\bf 338}, 635 (1989).
\bibitem{emery2} V.J. Emery and S.A. Kivelson, Physica C {\bf 209}, 597 (1993).
\bibitem{emery1} V.J. Emery, Phys. Rev. Lett. {\bf 58}, 2794 (1987).
\bibitem{dagotto} See, e.g., E. Dagotto (submitted to Rev. Mod. Phys.).
\bibitem{zhangrice} F.C. Zhang, and T.M. Rice, Phys. Rev. B
{\bf 37}, 3759 (1988).
\bibitem{shastry} B.S. Shastry, Phys. Rev. Lett. {\bf 63}, 1288 (1989).
\bibitem{goodingelser} R.J. Gooding and V. Elser, Phys. Rev. B
{\bf 41}, 2557 (1990); F.C. Zhang and T.M. Rice, {\em ibid} 2560 (1990).
\bibitem{schlueter} M. Schl\"uter (unpublished).
\bibitem{trugman} S.A. Trugman, Phys. Rev. B {\bf 37}, 1597 (1988). 
\bibitem{clarke} D.G. Clarke, Phys. Rev. B {\bf 48}, 7520 (1993); Ph.D. Thesis,
Princeton University, 1993 (unpublished).
\bibitem{stephan} W. Stephan and P. Horsch, Phys. Rev. Lett. {\bf 66},
2258 (1991).
\bibitem{gagliano} E. Gagliano, S. Bacci, and E. Dagotto, Phys. Rev. B
{\bf 42}, 6222 (1990).
\bibitem{leung} P.W.Leung, and P.E. Oppenheimer, Comp. in Phys. {\bf 6}, 603 (1992).
\bibitem{finitesize} This is important, since some finite--size effects 
were possibly present in Ref. \cite {GVL}. 
For example, one indication of these finite size effects 
is the band width of the one--hole band structure.
For the $\sqrt {18} \times \sqrt {32}$ cluster we obtain
a band width of .67 for $J=.4$ and $t=1$. This is very much consistent
with the expected band width, as may be seen via comparison
with \cite {poilblanc1}.
\bibitem{poilblanc1} D. Poilblanc, T.M. Ziman, H.J. Schulz, 
and E. Dagotto, Phys. Rev. B {\bf 47}, 14267 (1993).
\bibitem{24site} For the $\sqrt {18} \times \sqrt {32}$ 24--site cluster,
there is no $(\pi,0)$ state. Instead, the closest
wavevector to $(\pi,0)$ is $({5\pi\over 6}, {\pi\over 6})$, and it
is for this wave vector that we find the one--hole ground state
for the electron--doped system. See Table I.
\bibitem{wells} B.O. Wells, Z.--X. Shen, A. Matsuura, D.M. King, 
M.A. Kastner, and R.J. Birgeneau, submitted to Phys. Rev. Lett.
\bibitem{spiral} B.I. Shraiman and E.D. Siggia, Phys. Rev. Lett. {\bf 62}, 1564 (1989).
\bibitem{incomdirection} As discussed in Ref. \cite {GVL}, the direction
of the incommensurability that we find here is not in agreement with
experiment. This is to be expected, as our $\sqrt {8} \times \sqrt {32}$
cluster does not have many wave vectors along the important, e.g., 
$(\pi \pm \delta, \pi)$ direction.
\bibitem{noha} R.J. Gooding, N. Salem, and A. Mailhot, Phys. Rev. B {\bf 49}, 6067 (1994).
\bibitem{reiter} G.F. Reiter, Phys. Rev. B {\bf 49}, 1536 (1994).
\bibitem{unpublished1} R.J. Gooding (unpublished).
\bibitem{cho} J.H. Cho, F.C. Chou, and D.C. Johnston, Phys. Rev. Lett.
{\bf 47}, 222 (1993).
\bibitem{poilblanc2} D. Poilblanc, Phys. Rev. B {\bf 49}, 1477 (1994).
\bibitem{chen} K.J.E. Vos, P.W. Leung, L. Chen, and R.J. Gooding (in preparation).
\bibitem{riera} J.A. Riera and A.P. Young, Phys. Rev. B {\bf 39}, 9697 (1989).
\bibitem{chub123} K. Musaelian and A.V. Chubukov (unpublished).
\bibitem{liu} R. Liu, B.W. Veal, A.P. Paulikas, J.W. Downey, P.J. Kostic, S. Flescher,
U. Welp, C.G. Olson, X. Wu, A.J. Arko, and J.J. Joyce, Phys. Rev. B {\bf 46}, 11056 (1992).

\end{references}
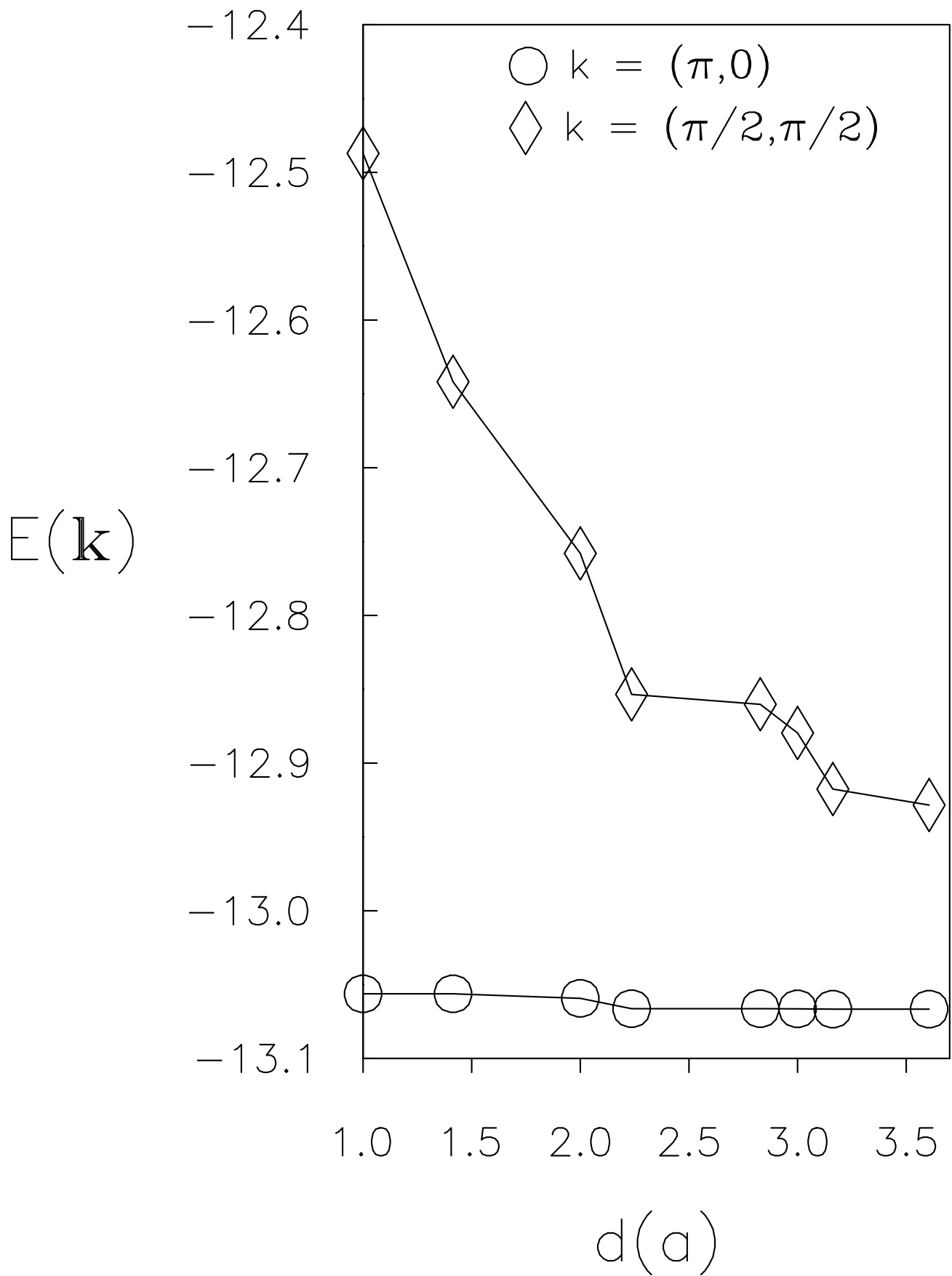
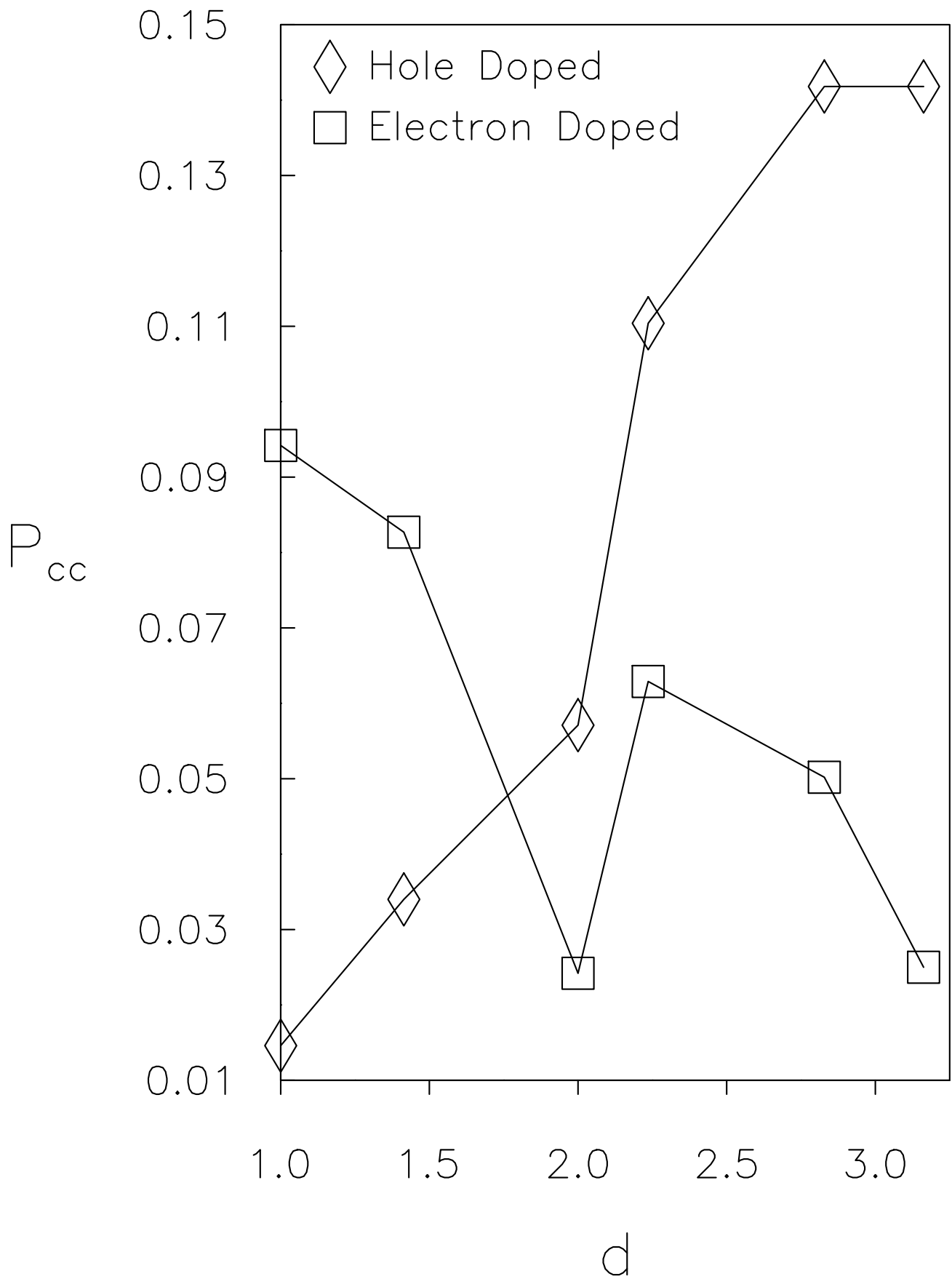
\end{document}